\documentclass[aps,showpacs,groupedaddress,pra,twocolumn,secnumarabic,nobalancelastpage,preprintnumbers]{revtex4}

\usepackage{graphics}
\usepackage[pdftex]{graphicx,color}
\usepackage{natbib}
\usepackage{amsmath}
\usepackage{amssymb}
\usepackage{braket}
\usepackage{bbm}
\usepackage[normalem]{ulem}
\newcommand{\newterm}[1]{\textit{#1}}

\newcommand{\im}{i}
\newcommand{\tr}{\textrm{Tr}}
\newcommand{\myD}{\mathcal{D}}
\newcommand{\sst}{\textrm{ss}}
\newcommand{\Stabs}{\mathcal{S}}
\newcommand{\myFid}{\mathcal{F}}

\begin{document}


\title{Stabilizing entanglement in the presence of local decay processes}

\author{Simeon Sauer}
\author{Clemens Gneiting}
\author{Andreas Buchleitner}
\affiliation{Physikalisches Institut, Albert-Ludwigs-Universit\"at Freiburg, Hermann-Herder-Stra{\ss}e~3, D-79104 Freiburg, Germany}
\date{\today}

\begin{abstract}
We investigate to what extent a suitably chosen system Hamiltonian can counteract local dissipative processes and preserve entanglement in the stationary state. The results determine prospects and limitations of dissipative state preparation schemes based on natural dissipative processes -- in contrast to engineered, typically non-local dissipative schemes.
As an exemplary case, we determine the stationary state of two spontaneously decaying two-level systems with optimal entanglement properties. The corresponding system Hamiltonian is derived, and its possible experimental implementation is discussed in detail. Finally, we discuss generalizations for $N$ qubits.
\end{abstract}

\preprint{\textsf{published in Phys.~Rev.~A~{89}, 022327 (2014)}}

\maketitle

\section{Introduction}\label{sec:Intro}

The preparation of a specific quantum state is a highly nontrivial task. If the quantum dynamics is purely Hamiltonian, it requires the control Hamiltonian to be conditioned on the (possibly unknown) initial state of the system in order to end up in the desired target state. An alternative route to prepare quantum states is thus to rely on dissipation, driving the system state into a unique fixed point of the dynamics. In many experimental setups this has been accomplished by cooling the system down to its ground state.

Dissipative state preparation schemes instrumentalize this approach in a systematic way. The idea is to engineer the coupling of an open quantum system to its environment such that any desired target state becomes the unique stationary state of the system dynamics. If one has set up these couplings properly, one then must simply wait: Any initial state is eventually attracted by the desired target state.

In principle, any pure state can be prepared dissipatively \cite{Wellens2000,Diehl:2008,Verstraete:2009,Kraus:2008}. The preparation of a pure, entangled state in a multipartite quantum system, however, requires the engineering of dissipative processes that jointly act on more than one party. While such a constructive scheme can be realized under very specific, highly engineered experimental conditions \cite{Krauter:2011,Kastoryano:2011,Wang:2010a,Stannigel:2012}, it poses insurmountable challenges for generic multicomponent quantum systems. This severely constrains the practicability of environment engineering for entanglement control.

In the present work, we therefore avoid engineered dissipative processes and rely, instead, on \textit{naturally} occurring incoherent processes to prepare entanglement. No elaborate experimental control over the environment will be required. This obvious benefit comes at a price: In general, natural dissipation is of \textit{local} nature, i.e., it acts individually on each party of a composite quantum system. Therefore, in contrast to engineered, nonlocal dissipation, it always acts adversely to entanglement. Under these circumstances, only the coherent part of the time evolution, i.e., the system Hamiltonian, can create entanglement. The aim of this work therefore is to investigate to what extent entanglement can then still be preserved in the stationary state. In particular, we seek to identify those system Hamiltonians which yield the most entangled stationary state for given (local) dissipative dynamics. Our agenda is hence to determine the prospects and limitations of entanglement control in the presence of local dissipation. 

The paper is organized as follows: In Sec.~\ref{sec:Theory}, we review the general concept of dissipative state preparation and discuss the consequences of resorting to strictly local dissipative processes. From Sec.~\ref{sec:Statistics} onward, we consider the most prominent example of a natural dissipative process, namely the spontaneous decay in two-level systems (qubits). First, we study typical stationary states by drawing from an ensemble of random Hamiltonians and compare them to the most entangled state among \emph{all conceivable} stationary states; the latter was derived by us previously \cite{Sauer:2013}.
In Sec.~\ref{sec:OptimalH}, we explicitly provide the Hamiltonian that leads to this optimal stationary state and discuss its experimental implementation. A generalization to the case of many qubits follows. Finally, we conclude in Sec.~\ref{sec:Conclusion}.

\section{Dissipative state preparation with local processes}\label{sec:Theory}

Throughout this paper, we consider open quantum systems that evolve under a master equation of Lindblad form \cite{Breuer:2007}:
\begin{equation}\label{eq:Lindbladeq}
\dot \rho = -\im [H,\rho] + \sum_{k} \myD_k(\rho) \quad\quad (\textrm{with } \hbar\equiv 1).
\end{equation}
The Hamiltonian $H$ governs the coherent part of the evolution of the quantum state $\rho$, whereas each $\myD_k(\rho)$ describes an incoherent process, defined through a Lindblad operator $L_k$ and a respective rate $\gamma_k$:
\begin{equation}\label{eq:Dissipator}
\myD_k(\rho) = \gamma_k \left[ L_k \rho L_k^\dagger - \frac{1}{2}\left(L_k^\dagger L_k \rho + \rho L_k^\dagger L_k \right)\right].
\end{equation}
Together, the incoherent terms define the \newterm{dissipator} $\myD(\rho)\equiv \sum_k \myD_k(\rho)$.

\subsection{Dissipative state preparation}

In a dissipative state preparation scheme \cite{Diehl:2008,Verstraete:2009}, Hamiltonian $H$ and dissipator $\myD$ are engineered such as to output the desired target state $\rho_\sst$ as the stationary solution of \eqref{eq:Lindbladeq}, i.e.,
\begin{equation}\label{eq:stationaritycond}
0 = -\im [H,\rho_\sst] + \myD(\rho_\sst).
\end{equation}
If, in addition, $\rho_\sst$ is the \emph{unique} stationary state, any initial state eventually evolves into this target state \cite{Schirmer:2010}. This defines, hence, a convenient preparation scheme for $\rho_\sst$.

If no restrictions are imposed on the Lindblad operators $L_k$ and the Hamiltonian $H$, there exsists a straightforward procedure to state $H$ and $\myD$ such that an arbitrary pure state $\rho=\ket{\psi}\bra{\psi}$ of an $N$ qubit quantum register results as the unique stationary state of $\eqref{eq:Lindbladeq}$ \cite{Kraus:2008}: First, one chooses the Hamiltonian such that $\ket{\psi}$ is an eigenstate,
\begin{equation}\label{eq:HpsiEpsi}
H \ket{\psi} = E \ket{\psi}, \qquad \textrm{ i.e.,} \qquad [H,\ket{\psi}\bra{\psi}] = 0.
\end{equation}
In a second step, one engineers the incoherent processes such that
\begin{equation}
L_k = U \sigma_-^{(k)} U^\dagger,
\end{equation}
where $\sigma_-^{(k)}$ denotes the operator $\sigma_-=\ket{0}\bra{1}$ of spontaneous decay of the $k$th qubit and $U$ is a unitary transformation defined by $\ket{\psi} = U \ket{0}^{\otimes N}$. The working principle of this scheme becomes apparent in a rotated reference frame, defined by $\tilde \rho = U^\dagger \rho U$. In this frame, $\ket{0}^{\otimes N}$ is a stationary state of \eqref{eq:Lindbladeq}, because it is an eigenstate of $\tilde H=U^\dagger H U$, and it is annihilated by any $\tilde L_k = U^\dagger L_k U = \sigma_-^{(k)}$. Moreover, $\ket{0}^{\otimes N}$ is the unique stationary state in the rotated frame \cite{Schirmer:2010}. Consequently, $\ket{\psi}=U \ket{0}^{\otimes N}$ is the unique stationary state in the original frame.

As a major difficulty of this scheme, however, the involved dissipative processes are not naturally given, but have to be designed artificially. In particular, an entangled target state $\ket{\psi}$ requires the engineering of nonlocal Lindblad operators \footnote{Since an entangled state $\ket{\psi}$ is not connected to $\ket{0}^{\otimes N}$ via a local unitary transformation, the transformation $U$ does not factorize w.r.t. to the different qubits. As a consequence, at least some of the Lindblad operators $L_k =  U \sigma_-^{(k)} U^\dagger$ act non-trivially on more than one subsystem, and hence describe non-local incoherent processes. In fact, the more sophisticated the desired kind of entanglement, the more Lindblad operators must be engineered in a non-local fashion, since all the entanglement properties of $\ket{\psi}$ are encoded in $U$. E.g., if one wants to entangle only two out of $N$ qubits, say $\ket{\psi}=(\ket{00}+\ket{11})/\sqrt{2}\otimes\ket{0}^{\otimes (N-2)}$, then only $L_1$ and $L_2$ are non-local, while the remaining $L_k$'s are just $\sigma_-^{(k)}$. On the other hand, if one wants to prepare a (linear or 2D) $N$-qubit cluster state, all $N$ Lindblad operators become non-local \cite{Kraus:2008}.}.
This is, in general, an exceedingly difficult task in realistic setups. And even if such engineering is achieved, additional, uncontrolled dissipative processes remain unavoidable. They will compete with the engineered processes and reduce the performance of the preparation scheme.

In view of these obstacles, we investigate here the potential of employing naturally occurring incoherent processes for preparing entangled states, instead of resorting to artificially engineered ones. In this approach, no experimental control over the incoherent dynamics is required; consequently, the natural incoherent processes do not compete with any expensively engineered ones, but are themselves an essential driving force of the preparation scheme.

\subsection{Local incoherent processes}

A generic property of most natural dissipative processes is their local nature. This implies that the corresponding Lindblad operators $L_k$ have a strictly local structure,
\begin{equation}\label{eq:locL}
L_{k} = \mathbbm{1} \otimes \dots \otimes \mathbbm{1} \otimes \ell \otimes \mathbbm{1} \otimes \dots \otimes \mathbbm{1},
\end{equation}
where $\ell$ acts only on the $k$th subsystem.

At first glance, the goal to prepare entangled target states with strictly local Lindblad operators seems counterintuitive: Local Lindblad operators necessarily tend to a \textit{decrease} of entanglement in time \cite{Carvalho:2004} -- so why should the stationary state exhibit finite entanglement? This objection is resolved by recognizing that, besides the incoherent part of the dynamics \eqref{eq:Lindbladeq}, there is also the coherent part, generated by the Hamiltonian $H$. If $H$ consists of strictly local terms as well, it is easy to see that the stationary state is not entangled, as shown in Appendix~\ref{sec:AppendixProofs}.\ref{ssec:AppendixProof1}. But if $H$ comprises nonlocal terms (i.e., interactions between the different local sites) this is not necessarily true. The subject of the present paper is, hence, to find the optimal $H$ such that the stationary state is entangled in the strongest possible way.

As a first result in this regard, we exclude the possibility to prepare \emph{pure} entangled target states with a local dissipator:

\newcommand{\propeins}{\paragraph*{} \textit{Be $\rho=\ket{\psi}\bra{\psi}$ a pure state of $N$ qubits that is stationary state under the master equation \eqref{eq:Lindbladeq}. If one of the Lindblad operators $L_{k}$ has the local structure \eqref{eq:locL}, then $\ket{\psi}$ is separable with respect to the $k$th qubit.}}
\propeins

A proof of this statement given in Appendix~\ref{sec:AppendixProofs}.\ref{ssec:AppendixProof2}. Note that the statement is independent of the presence of additional (possibly non-locally engineered) Lindblad operators.
It implies, in particular, separability of $\ket{\psi}$ with respect to \textit{every} bipartition, if there is at least one local $L_{k}$ for each qubit. Hence, under these circumstances, a stationary state $\rho$ cannot be pure and entangled at the same time. (If only one qubit is affected by dissipation, however, $\ket{\psi}$ can very well be entangled with respect to the remaining qubits, as explicitly shown in an example in Appendix~\ref{sec:AppendixExamplePSIent}.)
The above statement curbs, in conclusion, the expectation to prepare pure entangled target states. It makes, however, no statement about \textit{mixed}, stationary states. In fact, in the state space of all quantum states, there are weakly mixed states in the vicinity of any pure, maximally entangled state. One of the questions tackled in this work is whether such weakly mixed, highly entangled states can become stationary states under local dissipation, or whether there is a fundamental threshold that limits the entanglement of the accessible stationary states to a submaximal value.

Let us summarize our discussion so far: Dissipative state preparation schemes employ dissipation for entanglement creation, rather than considering it as adverse. To this end, usually both the incoherent and the coherent part of the dynamics (represented by the Lindblad operators $L_k$ and Hamiltonian $H$, respectively) are engineered in a non-local fashion.
As a result, both of them can act in favor of entanglement production. This way, any pure state can become the unique stationary state of the dynamics \eqref{eq:Lindbladeq}. In contrast, we focus on naturally occurring, \emph{local} incoherent processes, which therefore are necessarily detrimental for entanglement. These compete with the coherent dynamics, which may include interactions between the local sites and can thus build up entanglement. This results in a stationary state that is either mixed or separable (with respect to any site that undergoes a local incoherent process).
However, the incoherent part of the time evolution has a second important role for dissipative state preparation, which remains untouched: It leads to a \textit{unique} stationary state, ensuring that any initial condition eventually evolves into the target state. This task cannot be accomplished by coherent dynamics alone.

\section{Preserving entanglement in the presence of spontaneous decay}\label{sec:Statistics}

In the following section, we focus on the most prominent example of a naturally occurring incoherent process: the omnipresent spontaneous decay of two-level systems with a decay rate $\gamma$. Thus, we consider two qubits, with a single Lindblad operator per site:
\begin{equation}\label{eq:Lsminus}
L_{1}= \sigma_- \otimes \mathbbm{1} , \quad L_{2}=  \mathbbm{1}\otimes\sigma_-.
\end{equation}
 In this case, the stationary state $\rho_\sst$ of the dynamics is unique, irrespectively of the Hamiltonian $H$ \cite{Schirmer:2010}. Nevertheless, $\rho_\sst$ itself depends on the particular choice of $H$. The task is hence to find the $H$ that results in the optimally entangled $\rho_\sst$.

There are several ways to quantify entanglement between two qubits. One possibility is {to employ an entanglement measure, such as the} \newterm{concurrence} $\mathcal{C}(\rho)$ \cite{Hill:1997}, which takes values between zero (for separable states) and one (for maximally entangled states).
If one is not only interested in high values of entanglement alone, but rather wants to specifically create one of the four maximally entangled Bell states,
\begin{eqnarray}
\ket{\Phi_\pm}&=&\frac{1}{\sqrt 2}(\ket{11}\pm\ket{00}), \label{eq:BelldefPhi} \\
\ket{\Psi_\pm}&=&\frac{1}{\sqrt 2}(\ket{10}\pm\ket{01}), \label{eq:BelldefPsi}
\end{eqnarray}
the \newterm{Bell state fidelities}
\begin{eqnarray}\label{eq:Fiddef}
	\myFid_{\Phi_\pm}(\rho) & \equiv & \braket{\Phi_\pm | \rho | \Phi_\pm} , \\
	\myFid_{\Psi_\pm}(\rho) & \equiv & \braket{\Psi_\pm | \rho | \Psi_\pm} ,
\end{eqnarray}
are alternative quantities of interest, ranging likewise from zero to one. Their value is directly related to the fidelity of a teleportation protocol via the state $\rho$ \cite{Horodecki:1999}. We will consider both, concurrence and fidelities, as entanglement quantifiers in the following.

\subsection{Introductory examples: Ising and Heisenberg interaction of two qubits}\label{ssec:Ising}

To gain first insight on how much entanglement can be preserved in the stationary state in the presence of spontaneous decay, we consider the exemplary Hamiltonian of two qubits with Ising interaction \cite{Amico:2008},
\begin{equation}\label{eq:HIsing}
H= \frac{\Delta}{2} \, (\mathbbm{1} \otimes \sigma_z + \sigma_z \otimes \mathbbm{1} ) + J \, \sigma_x \otimes \sigma_x,
\end{equation}
with local energy splitting $\Delta$ and interaction strength $J$.
The stationarity condition \eqref{eq:stationaritycond} can in this case be solved explicitly for the stationary state $\rho_\sst$ \cite{Guerreschi:2012}:
\begin{eqnarray}\label{eq:rhoIsing}\nonumber
\rho_\sst = \frac{1}{1 + |x|^2} \left( \frac{\mathbbm{1}_4}{4} + |x|^2 \ket{00}\bra{00}  \right. \\ \left. + \frac{x}{2} \ket{00}\bra{11} + \frac{x^*}{2} \ket{11}\bra{00} \right),
\end{eqnarray}
with $x \equiv (\Delta + i \gamma)/J$.
Since this state is of ``X form'' \cite{Yu:2007}, there is a simple expression for its concurrence:
\begin{equation}\label{eq:Isingresult}
\mathcal{C}(\rho_\sst) = \max \left\{ 0, \frac{ |x| - \frac{1}{2}}{1 + |x|^2} \right\}.
\end{equation}
For the Bell state fidelities, one obtains
\begin{eqnarray} \nonumber
	\myFid_{\Phi_\pm}(\rho_\sst) &= &\frac{1}{2}+\frac{\pm2 \,\mathrm{Re}(x)-1}{4(1+|x|^2)}, \\
	\myFid_{\Psi_\pm}(\rho_\sst) &= &\frac{1}{4(1+|x|^2)}.  \label{eq:IsingresultFid}
\end{eqnarray}
In the limit of strong interaction [$J \gg \max (\Delta, \gamma)$, i.e., $|x| \ll1$], the stationary state approaches the completely mixed state $\mathbbm{1}_4 / 4$, which is not entangled and has poor Bell state fidelities $\myFid_{\Phi_\pm}(\rho_\sst)=\myFid_{\Psi_\pm}(\rho_\sst)=\frac{1}{4}$. In the opposite case of weak interaction [$J \ll \max (\Delta, \gamma)$, i.e., $|x| \gg 1$], we have $\rho_\sst = \ket{00}\bra{00}$, which is not entangled, either, with Bell state fidelities $\myFid_{\Phi_\pm}(\rho_\sst)=\frac{1}{2}$ and $\myFid_{\Psi_\pm}(\rho_\sst)=0$.

Maximizing expression \eqref{eq:Isingresult} leads to an optimal concurrence value of $\mathcal{C}(\rho_\sst) = (\sqrt{5}-1)/4\approx 0.31$, which is reached at the golden ratio $|x|=(1+\sqrt{5})/2$, i.e., at $\sqrt{\Delta^2+\gamma^2}/J = (1+\sqrt{5})/2$. Likewise, the optimal fidelities are $\myFid_{\Phi_\pm}(\rho_\sst)=(3+\sqrt{5})/8\approx 0.65$ for $x=\pm(1+\sqrt{5})/2$, and $\myFid_{\Psi_\pm}(\rho_\sst)=\frac{1}{4}$ for $x=0$.
Fig.~\ref{fig:varygamma} shows the concurrence and the Bell state fidelities as a function of $J/\gamma$, for $\Delta=0$ and $\Delta=J$ (solid and dashed red curves, respectively; shaded data refer to Sec. \ref{ssec:MCsampling}).

It is interesting to repeat this analysis for the XXZ Heisenberg interaction \cite{Amico:2008}, i.e., for
\begin{equation}\label{eq:HHeisenb}
H= \frac{\Delta}{2} \, (\mathbbm{1} \otimes \sigma_z + \sigma_z \otimes \mathbbm{1} ) + J \, (\sigma_x \otimes \sigma_x + \sigma_y \otimes \sigma_y + \alpha \sigma_z \otimes \sigma_z).
\end{equation}
For any choice of the anisotropy factor $\alpha$, the steady state of the master equation \eqref{eq:Lindbladeq} is then simply the de-excited state $\ket{00}\bra{00}$. This is because the deexcited state is annihilated by both Lindblad operators $L_1$ and $L_2$ from Eq.~\eqref{eq:Lsminus}, and it commutes with the Hamiltonian \eqref{eq:HHeisenb} \footnote{Note that this argument remains also valid for a linear spin chain of $N>2$ qubits, so that the steady state is always the de-excited state $\ket{0}^{\otimes N}\bra{0}^{\otimes N}$.}.  Due to the fact that the stationary state is unique with the given Lindblad operators, and that the deexcited state is obviously not entangled, we conclude that the XXZ Heisenberg interaction never leads to stationary entanglement.

In summary, for two qubits with the Ising Hamiltonian \eqref{eq:Lsminus} and local spontaneous decay \eqref{eq:HIsing}, it is not possible to achieve a stationary state that exceeds the values of $\mathcal{C}(\rho_\sst) \approx 0.31$, $\myFid_{\Phi_\pm}(\rho_\sst) \approx 0.65$ and $\myFid_{\Psi_\pm}(\rho_\sst)=\frac{1}{4}$, no matter how one adjusts the parameters $\Delta$ and $J$ of the Hamiltonian in comparison to the dissipation rate $\gamma$. The Heisenberg interaction \eqref{eq:HHeisenb}, on the other hand, leads to no stationary entanglement at all. In the following, we investigate systematically whether there are Hamiltonians which perform better than that.

\subsection{Random Hamiltonians}\label{ssec:MCsampling}

As a first step towards a systematic search for the optimal Hamiltonian, we study an ensemble of random Hamiltonians $H$ and investigate the distribution of the concurrence of the resulting stationary states. This way, we explore the entanglement properties of \emph{typical} stationary states. The Hamiltonian is drawn from the Gaussian Unitary Ensemble (GUE) \cite{Mehta:2004}; i.e., we set $H=J (X+iY)/2+h.c.$, where $X$ and $Y$ are real $4 \times 4$ matrices with random entries that are independently drawn from a standard normal distribution. The parameter $J$ thus defines the energy scale of the Hamiltonian. However, since the stationarity condition \eqref{eq:stationaritycond} is independent of a scaling factor, the only parameter that determines the stationary state is the ratio $J/\gamma$, describing the relative strength of coherent and dissipative dynamics.

\begin{figure*}[tb]
\begin{tabular}{ccc}
	(a) & (b) & (c) \\
	\includegraphics[width=.33\textwidth]{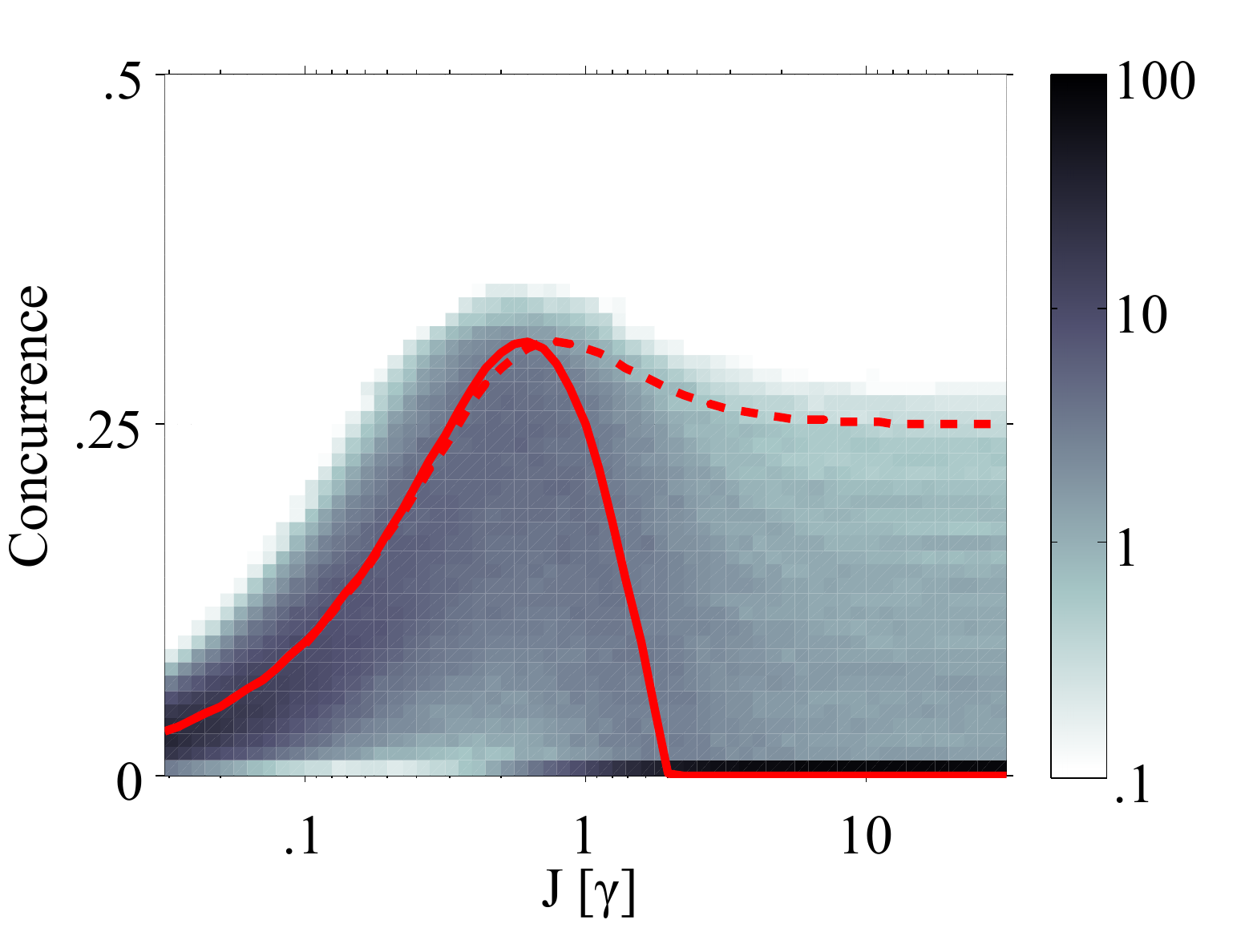} &
	\includegraphics[width=.33\textwidth]{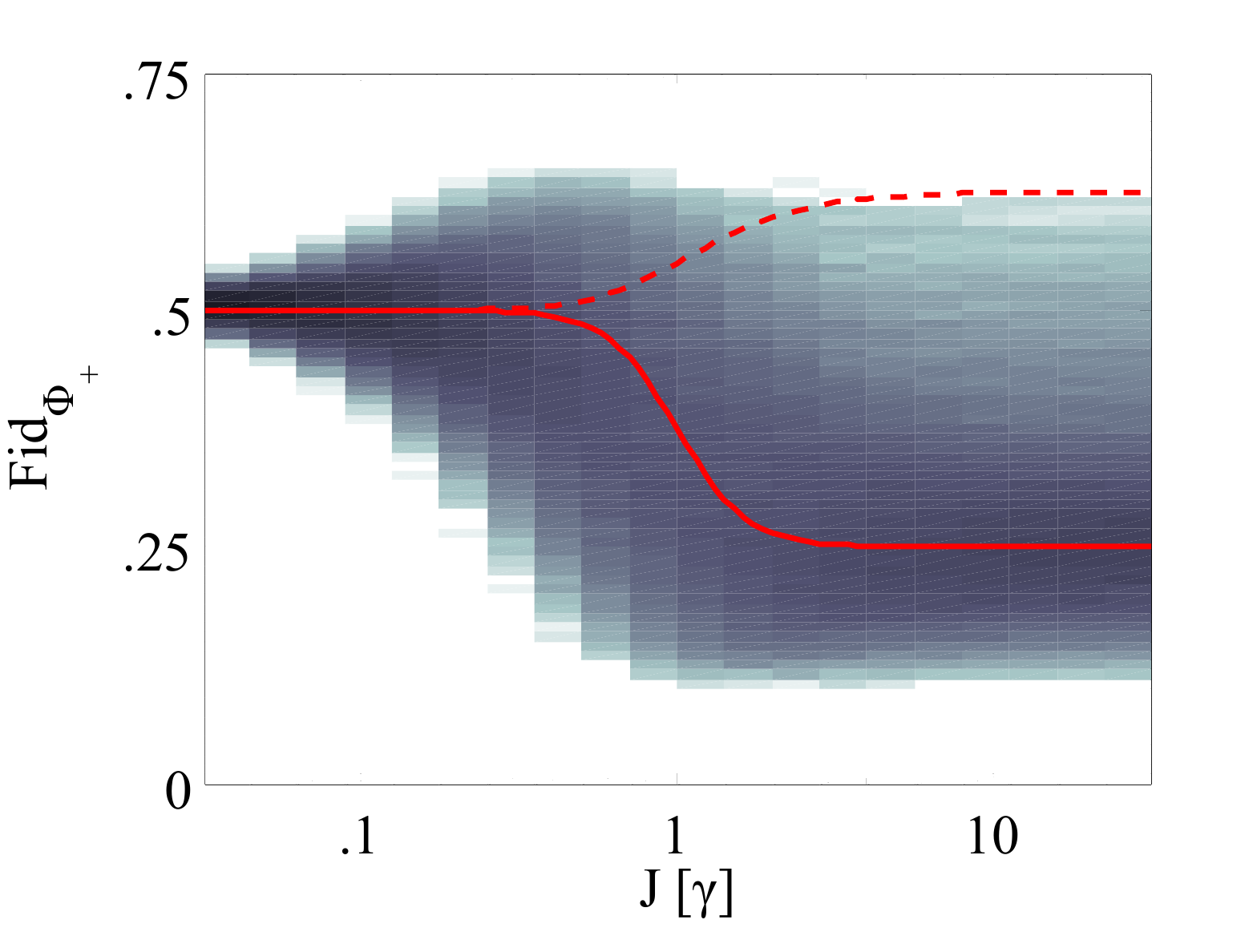} &
	\includegraphics[width=.33\textwidth]{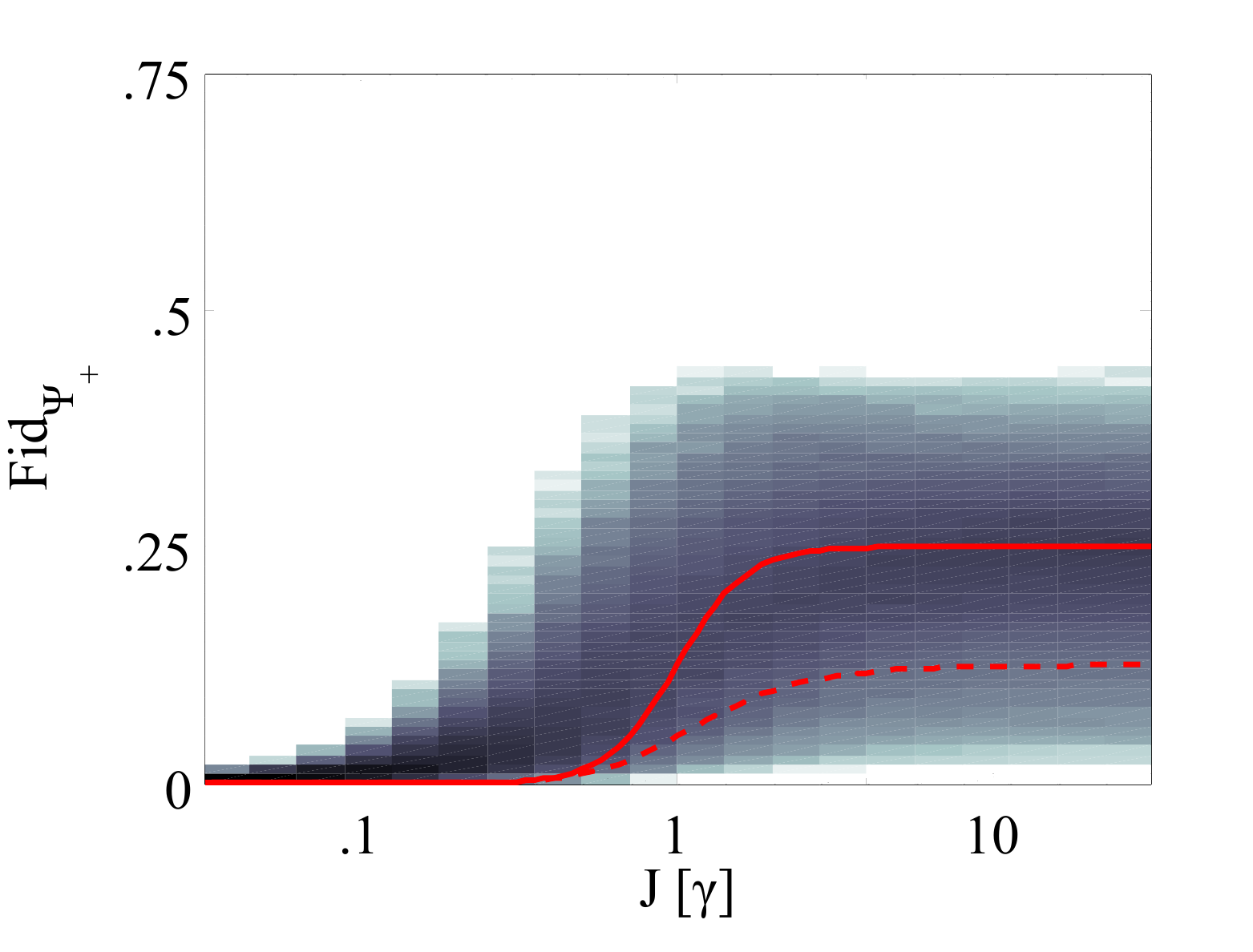}
\end{tabular}
\caption{(Color online) Distribution of (a) the concurrence $\mathcal{C}(\rho_\sst)$, and [(b), (c)] the Bell state fidelities $\myFid_{\Phi_+}(\rho_\sst)$ and $\myFid_{\Psi_+}(\rho_\sst)$, Eq. \eqref{eq:Fiddef}, of the stationary state $\rho_\sst$. The distributions were obtained for the case of two qubits, the coherent dynamics of which is generated by an ensemble of $10^4$ Hamiltonians having random entries with zero mean and standard deviation $J$, and the incoherent dynamics by spontaneous decay at rate $\gamma$. On the abscissa, the ratio $J/\gamma$ varies over three orders of magnitude. At each value of $J/\gamma$, the color code indicates, on a logarithmic scale, the probability density to find a certain concurrence or fidelity value (ordinate) in the ensemble. While the concurrence can in principle reach values up to $\mathcal{C}=1$, the probability to find values $\mathcal{C}> 0.35$ vanishes. (We therefore plot only the interval $0 \le \mathcal{C} \le 0.5$.) Likewise, the Bell state fidelities are limited to $\myFid_{\Phi_+} \lessapprox 0.65$ and $\myFid_{\Psi_+}\lessapprox 0.5$. The red curves visualize the results \eqref{eq:Isingresult} and \eqref{eq:IsingresultFid} obtained for the Ising Hamiltonian \eqref{eq:HIsing} with $\Delta=0$ (solid curves), and $\Delta=J$ (dashed curves). Figures (b) and (c) are identical if one plots $\myFid_{\Phi_-}$ and $\myFid_{\Psi_-}$ instead of $\myFid_{\Phi_+}$ and $\myFid_{\Psi_+}$, with the only difference being that the dashed red curve corresponds to $\Delta=-J$ in that case.}\label{fig:varygamma}
\end{figure*}

For different values of $J/\gamma$, we generated $10^4$ realizations of $H$. For each $H$, we obtained the corresponding stationary state $\rho_\sst$ by numerically solving \eqref{eq:stationaritycond} and calculated the concurrence $\mathcal{C}(\rho_\sst)$ and the fidelities $\mathrm{Fid}_{\Phi_\pm}(\rho_\sst)$ and $\mathrm{Fid}_{\Psi_\pm}(\rho_\sst)$. Fig.~\ref{fig:varygamma} shows the results as a function of $J/\gamma$ on the abscissa. The gray scale indicates, on a logarithmic scale, the probability density to find a certain value of the quantity of interest in the ensemble. Hence, each column of the plots represents a histogram of the concurrence or of the fidelity at a fixed value of $J/\gamma$. 

The solid and dashed red curves visualize the analytical results \eqref{eq:Isingresult} and \eqref{eq:IsingresultFid} obtained for the Ising Hamiltonian \eqref{eq:HIsing}. Regarding its ability to generate entangled stationary states, this Hamiltonian appears as a generic member of the GUE. E.g., the statistical ensemble achieves the highest concurrence value of $\mathcal{C}(\rho_\sst)\approx 0.35$ when coherent and dissipative dynamics are of comparable strength, i.e., around $J/\gamma\approx 1$ -- similar to the Ising Hamiltonian, which reaches $C(\rho_\sst)=(\sqrt{5}-1)/4 \approx 0.31$ at $J/\gamma = 2/(1+\sqrt{5})\approx 0.61$ (for $\Delta=0$).
For weak coherent dynamics, $J/\gamma \ll 1$, the stationary state approaches the deexcited state $\rho_\sst=\ket{00}\bra{00}$, independently of the Hamiltonian. Hence, the concurrence of both the statistical ensemble and the Ising Hamiltonian vanish in this limit, whereas the Bell state fidelities approach $\myFid_{\Phi_+}=\frac{1}{2}$ and $\myFid_{\Psi_+}=0$.

The most important conclusion to be drawn from Fig.~\ref{fig:varygamma}, {however,} is that typical stationary states yield concurrence $\mathcal{C}>0.35$ with vanishing probability, irrespectively of $J/\gamma$. The value of $\mathcal{C}\approx 0.31$ derived for the Ising Hamiltonian is therefore close to the maximal entanglement that can be expected for a generic stationary state. The same holds for the $\Phi_\pm$-fidelity: The random ensemble does not exceed the threshold $\myFid_{\Phi_ \pm}\approx 0.65$ of the Ising Hamiltonian. Only for the ${\Psi_ \pm}$-fidelity, the ensemble significantly outperforms the Ising Hamiltonian and reaches up to $\myFid_{\Psi_ \pm}\approx 0.5$.

\subsection{The optimally entangled stationary state}\label{ssec:Derivrhoopt}

The foregoing statistical investigation of typical stationary states cannot exclude the existence of atypical stationary states with better entanglement properties. Therefore, we derive in the following the {most} entangled state $\rho^{*}$ among \emph{all conceivable} stationary states that can emerge in the presence of spontaneous decay.
This is achieved with a general method to {solve optimization problems for the stationary state of open quantum systems}, which was presented in \cite{Sauer:2013}. In the following, we briefly recapitulate the main ideas {of this method}.

The standard procedure to find the stationary state $\rho^{*}$ that {maximizes} an objective function $O(\rho)$ -- in our case the concurrence or the Bell state fidelity -- for given dissipative dynamics (here, spontaneous decay) requires two steps: First, one inverts the stationarity condition \eqref{eq:stationaritycond}, such that the stationary state becomes a function of the Hamiltonian, $\rho(H)$. Second, one optimizes the objective $O(\rho(H))$ over all conceivable Hamiltonians $H$.
This procedure, however, has two drawbacks: First, $\eqref{eq:stationaritycond}$ can in general only be inverted by numerical means. Second, the set of Hamiltonians is unbounded, rendering the maximization difficult.

Therefore, we {have developed} a different method to tackle such optimization problems \cite{Sauer:2013}. The core idea is to optimize the quantity of interest $O(\rho)$ over the set of \newterm{stabilizable states} $\Stabs$ \cite{Recht:}, defined as
\begin{equation}
\Stabs \equiv \{ \rho \in \mathcal{Q} \ | \ \exists H: 0 = -i[H,\rho]+\myD(\rho)\}.
\end{equation}
($\mathcal{Q}$ denotes the set of quantum states.) By definition, $\Stabs$ contains all quantum states that can become stationary under given dissipative dynamics $\myD(\rho)$. Every state $\rho\in\Stabs$ corresponds to a suitable Hamiltonian $H$ that renders this particular state stationary. The set of stabilizable states $\Stabs$ itself, however, does not depend on the Hamiltonian, but is exclusively determined by the dissipator $\myD(\rho)$.

As an advantage of this approach, the set $\Stabs$, being a subset of the state space $\mathcal{Q}$, is a bounded set. This facilitates the optimization procedure. Moreover, for a given state $\rho\in\Stabs$, it is possible to solve the stationarity condition \eqref{eq:stationaritycond} for the corresponding Hamiltonian $H$ in a systematic way: Taking the spectral decomposition $\rho=\sum_\alpha p_\alpha \ket{\alpha}\bra{\alpha}$, and ``sandwiching'' Eq. \eqref{eq:stationaritycond} with eigenstates $\bra{\alpha}$ from the left and $\ket{\beta}$ from the right, one arrives at
\begin{equation}
	 0 = -i (p_\beta - p_\alpha) \braket{\alpha | H| \beta} + \braket{\alpha | \myD(\rho) | \beta}.
\end{equation}
For $p_\alpha \ne p_\beta$, this leads to
\begin{equation}
	  \braket{\alpha | H| \beta} = \frac{i \bra{\alpha} \myD(\rho) \ket{\beta}  } { p_\alpha - p_\beta }.
\end{equation}
On the other hand, if $p_\alpha = p_\beta$, the dissipative matrix element $\bra{\alpha} \myD(\rho) \ket{\beta}$ must vanish, implying that the Hamiltonian matrix element $\braket{\alpha | H| \beta}$ can be chosen arbitrarily. Note that this is in particular the case for the diagonal elements $\braket{\alpha | H| \alpha}$. Hence, given a stabilizable state $\rho\in \Stabs$, the corresponding Hamiltonian is
\begin{equation}\label{eq:Hcompensation}
	H = \sum_{p_\alpha \ne p_\beta} \frac{i \bra{\alpha} \myD(\rho) \ket{\beta}  } { p_\alpha - p_\beta } \ket{\alpha}\bra{\beta}+ \sum_{p_\alpha = p_\beta}x_{\alpha\beta}\ket{\alpha}\bra{\beta},
\end{equation}
with arbitrary elements $x_{\alpha\beta}$ (fulfilling $x_{\alpha\beta}=x_{\beta\alpha}^*$).

To characterize the set of stabilizable states $\Stabs$, one can exploit the fact that the coherent part of the evolution, generated by $-i[H,\rho]$, induces strictly unitary dynamics, which leaves the spectrum $\{p_\alpha \}$ of $\rho$ invariant. Thus, only the dissipative term $\myD(\rho)$ can alter the spectrum of $\rho$. At a stationary state $\rho_\sst$, however, we have $\myD(\rho_\sst)=i[H,\rho_\sst]$, implying that the dissipative dynamics compensates for the coherent evolution. Thus, $\myD(\rho_\sst)$ merely induces unitary dynamics as well (at the particular state $\rho_\sst$), and the evolution under $\myD(\rho_\sst)$ alone must leave the spectrum $\{p_\alpha \}$ of $\rho_\sst$ invariant. The spectrum, in turn, is uniquely defined by its leading $d$ statistical moments
\begin{equation}
\mu_n\equiv\sum_{\alpha=1}^d (p_\alpha)^n = \tr (\rho^n),
\end{equation}
where $d$ refers to the dimension of the quantum state $\rho$.
This implies that the evolution under $\myD(\rho)$ is unitary in the neighborhood of $\rho$ if and only if it leaves all moments $\mu_n$ of $\rho$ invariant, i.e., if and only if
\begin{equation}
0=  \frac{d}{dt} \mu_n |_{H=0}=\frac{d}{dt} \tr (\rho^n) |_{H=0} \overset{\eqref{eq:Lindbladeq}}{=} n \tr [\rho^{n-1} \myD(\rho)]
\end{equation}
holds for $n=2,\dots,d$. ($n=1$ is omitted, because $\mu_1=\tr(\rho)=1$ is always conserved.)
This defines $d-1$ necessary criteria for $\rho\in\Stabs$:
\begin{equation}\label{eq:conservespectrum}
\rho \in \Stabs \quad \Rightarrow \quad \forall n \in \{2,...,d\}: \ \tr [\rho^{n-1} \myD(\rho)].
\end{equation}
If $\rho$ has non-degenerate eigenvalues, these criteria (taken together) are also sufficient for $\rho\in\Stabs$ \cite{Sauer:2013}.
Ultimately, these arguments ensure that the set $\Stabs$, which contains all accessible stationary states for a given dissipator $\myD(\rho)$, is generated by collecting all $\rho$ that obey condition \eqref{eq:conservespectrum} \footnote{Strictly speaking, condition \eqref{eq:conservespectrum} does not apply for degenerate states. These, however, do not cause severe problems: Suppose that a degenerate state $\rho$ obeys condition \eqref{eq:conservespectrum}, but does not lie in $\Stabs$. For a reasonably well-behaved dissipator $\myD(\rho)$, there will be a state $\rho'$ in the vicinity of $\rho$ that also fulfills \eqref{eq:conservespectrum} and has slightly different, non-degenerate eigenvalues, so that $\rho'$ lies in $\Stabs$. (Note, however, that the Hamiltonian rendering $\rho'$ stationary will become unboundedly large as $\rho'$ approaches $\rho$, since it has \textit{almost} degenerate eigenvalues, rendering the denominator in \eqref{eq:Hcompensation} small.)}.

For our purposes, it is convenient to reexpress criterion \eqref{eq:conservespectrum} in terms of the generalized, 15-dimensional Bloch vector $\vec{\mathfrak{r}}_\rho$ that represents the quantum state $\rho$ of two qubits, as defined in Appendix~\ref{sec:AppendixQuadricN2}.
In this representation, the set of stabilizable states $\Stabs$ is given by the intersection of three nonlinear hypersurfaces $\Stabs^{(n)}$, each representing one of the constraints imposed by condition \eqref{eq:conservespectrum}. The lowest order constraint ($n=2$) is quadratic in the Bloch vector:
\begin{equation}\label{eq:quadric}
 \vec{\mathfrak{r}}_\rho \cdot ( \mathfrak{D}\vec{\mathfrak{r}}_\rho + \vec{\mathfrak{c}} ) = 0.
\end{equation}
(The entries of the constant matrix $\mathfrak{D}$ and the vector $\vec{\mathfrak{c}}$ are given in Appendix~\ref{sec:AppendixQuadricN2}.)
The higher order constraints for $n=3,4$ lead to polynomial expressions of third and fourth degree in the Bloch vector, resulting in hypersurfaces $\Stabs^{(3)}$ and $\Stabs^{(4)}$, which we refrain {from analyzing}.
Instead, in order to determine the optimal stationary state, we proceed as follows: First, we determine the most entangled state in $\Stabs^{(2)}$, i.e., among all those states that fulfill constraint \eqref{eq:conservespectrum} for $n=2$. Then, we verify that the resulting optimal state $\rho^{*}\in\Stabs^{(2)}$ lies in $\Stabs$. If this is the case, it must also be the most entangled state in $\Stabs$, since $\rho^{*} \in \Stabs^{(2)}$ is a \textit{necessary} condition for $\rho^{*} \in \Stabs$, i.e., $\Stabs \subset \Stabs^{(2)}$. Thus, if the most entangled state in $\Stabs^{(2)}$ turned out not to lie in $\Stabs$, the procedure would still provide an upper bound for the maximal entanglement in $\Stabs$.

The maximization of the Bell state fidelities $\myFid_{\Phi_\pm}(\rho)$ and $\myFid_{\Phi_\pm}(\rho)$ over all  $\rho\in\Stabs^{(2)}$ can be carried out analytically, since the latter are linear quantities in $\rho$. In Bloch notation, the objective function becomes
\begin{equation}
\myFid_{X}(\vec{\mathfrak{r}}_\rho) = \vec{\mathfrak{r}}_{X} \cdot \vec{\mathfrak{r}}_\rho,
\end{equation}
where $X$ refers to either the $\ket{\Phi_\pm}$ or the $\ket{\Psi_\pm}$ Bell state.
Applying Lagrange's method with a multiplier $\lambda$, one finds that the optimal $\vec{\mathfrak{r}}_{\rho^{*}}$ must satisfy
\begin{eqnarray} \nonumber
 0  & = &\left. \vec\nabla_{ \vec{\mathfrak{r}}_\rho} \left[  \vec{\mathfrak{r}}_X \cdot \vec{\mathfrak{r}}_\rho - \lambda \vec{\mathfrak{r}}_\rho \cdot  ( \mathfrak{D}\vec{\mathfrak{r}}_\rho + \vec{\mathfrak{c}}) \right]  \right|_{\vec{\mathfrak{r}}_{\rho} = \vec{\mathfrak{r}}_{\rho^{*}}} \\ \label{eq:Lagrange1}
\Rightarrow  0 & = &  \vec{\mathfrak{r}}_X - \lambda  ( \mathfrak{D}\vec{\mathfrak{r}}_{\rho^{*}} + \vec{\mathfrak{c}}),  \\
\textrm{and }  0 & = & \vec{\mathfrak{r}}_{\rho^{*}} \cdot  (\mathfrak{D}\vec{\mathfrak{r}}_{\rho^{*}} + \vec{\mathfrak{c}} ) \label{eq:Lagrange2}.
\end{eqnarray}

For the $\ket{\Psi_\pm}$ Bell fidelity, the density matrix corresponding to the solution $\vec{\mathfrak{r}}_{\rho^{*}}$ of this system of {16 equations} is
\begin{equation}\label{eq:rhoopti}
\rho^{*}_{\Psi_\pm} = \frac{1}{2} \ket{00}\bra{00} +  \frac{1}{2} \ket{\Psi_\pm}\bra{\Psi_\pm},
\end{equation}
yielding $\myFid_{\Psi_\pm}(\rho^{*}_{\Psi_\pm})=\frac{1}{2}$. This state lies indeed in $\Stabs$, as shown in the following section. Anticipating this result, we have proven that the upper bound of $\myFid_{\Psi_\pm}\lessapprox 0.5$ that is observed for typical stationary states in Fig.~\ref{fig:varygamma}(b), marks indeed the optimal value among \textit{all} accessible stationary states.

Maximizing instead the fidelity with the $\Phi_\pm$ Bell states, the solution of \eqref{eq:Lagrange1} and \eqref{eq:Lagrange2} leads to
\begin{eqnarray}\label{eq:rhooptipsi}
\rho^{*}_{\Phi_\pm} = \frac{1}{12} \left(\mathbbm{1}_4 + 9 \ket{00}\bra{00} - \ket{11}\bra{11} \right. \nonumber \\ 
\left. \pm 3 \ket{00}\bra{11} \pm 3 \ket{11}\bra{00}\right),
\end{eqnarray}
which yields $\myFid_{\Phi_\pm}(\rho^{*})=\frac{2}{3}$. This solution, however, does not describe a valid quantum state, since its smallest eigenvalue is $(5 - \sqrt{34})/12 \approx -0.07$. Hence, this analysis only provides an upper bound of $\frac{2}{3}$ for the true optimal value of $\myFid_{\Phi_\pm}$ in $\Stabs$. Nevertheless, we have already encountered a valid quantum state that almost perfectly saturates this upper bound: In Sec. \ref{ssec:Ising}, we found $\myFid_{\Phi_\pm}(\rho_\sst)=(3+\sqrt{5})/8\approx 0.65$ for the stationary state \eqref{eq:Isingresult} [with $x=\pm(1+\sqrt{5})/2$]. This state is similar to the unphysical state \eqref{eq:rhooptipsi}, and we therefore conjecture that it is the true optimal stationary state for the ${\Phi_\pm}$-fidelity in $\Stabs$.

The concurrence $\mathcal{C}(\rho)$ cannot be optimized in the same analytical fashion, because it is not a linear function of $\rho$, rendering the analytical evaluation of the gradient in \eqref{eq:Lagrange1} intractable. However, high Bell state fidelity typically corresponds to strong entanglement. It is therefore reasonable to look at the concurrence of the fidelity-optimized states derived above. The ${\Phi_\pm}$-optimal state \eqref{eq:rhoIsing} (for $x=\pm(1+\sqrt{5})/2$) yields $\mathcal{C}=(\sqrt{5}-1)/4\approx 0.31$. The ${\Psi_\pm}$-optimal state \eqref{eq:rhoopti}, on the other hand, reaches $\mathcal{C}=\frac{1}{2}$. This exceeds significantly the upper bound of $\mathcal{C}\approx0.35$ that we observed for typical stationary states in Fig.~\ref{fig:varygamma}(a). Moreover, an optimization of the concurrence with numerical means does not improve on $\mathcal{C}=\frac{1}{2}$.
This strongly indicates that \eqref{eq:rhoopti} is optimal with respect to both $\myFid_{\Psi_\pm}$ \textrm{and} the concurrence. We therefore study in detail its preparation in the following section.

We emphasize that the optimal state \eqref{eq:rhoopti} has been discussed before in Refs.~\cite{Recht:,Sauer:2013}. However, its role as the \textit{optimal} among all stabilizable states has not yet been recognized in \cite{Recht:}, but was only discussed in our earlier work \cite{Sauer:2013}. In the latter work, on the other hand, we did not discuss its derivation and preparation in detail, but merely used it as an exemplary application of the general method developed there for the optimization of stationary states.

\section{The optimal Hamiltonian for spontaneous decay}\label{sec:OptimalH}

\subsection{The optimal Hamiltonian for two qubits}\label{ssec:OptimalHN2}

So far, we have not verified that $\rho^{*}_{\Psi_\pm}$ of Eq.~\eqref{eq:rhoopti} is indeed an accessible stationary state, i.e., that $\rho^{*}_{\Psi_\pm} \in \Stabs$. We prove this in the following by explicitly providing the Hamiltonian $H_\pm^*$ that renders $\rho^{*}_{\Psi_\pm}$ stationary under the master equation \eqref{eq:Lindbladeq}, assuming spontaneous decay for both qubits at rate $\gamma$.

Since $\rho^{*}_{\Psi_\pm}$ has degenerate eigenvalues $\{\frac{1}{2},\frac{1}{2},0,0\}$, prescription \eqref{eq:Hcompensation} cannot be used to determine $H_\pm^*$. One may, however, consider the nearby, non-degenerate state
\begin{eqnarray}
\rho_{\epsilon} &=& \frac{1}{1+ \epsilon^2}\left(\rho^{*}_{\Psi_\pm} + \frac{3\epsilon^2}{2}\ket{00}\bra{00} - \frac{\epsilon^2}{2}\ket{\Psi_\pm}\bra{\Psi_\pm} \right. \nonumber\\
& & \left. \mp \epsilon\sqrt{1-\epsilon^2} \left(\ket{00}\bra{\Psi_\pm}+\ket{\Psi_\pm}\bra{00}\right) \right)
\end{eqnarray}
instead (with $\epsilon\ll1$), which also fulfills constraint \eqref{eq:conservespectrum} for $n=2$ and coincides with the state of interest $\rho^{*}_{\Psi_\pm}$ for $\epsilon=0$. Taking the corresponding Hamiltonian \eqref{eq:Hcompensation} for this state and performing the limit $\epsilon\rightarrow 0$, one arrives at the desired Hamiltonian that renders $\rho^{*}_{\Psi_\pm}$ stationary. It reads
\begin{eqnarray} \label{eq:HoptN2} \nonumber
H^{*}_{\pm} &=& \mathbbm{1} \otimes \left( \frac{\Delta}{2} \sigma_z + \frac{F}{2} \sigma_x \right) + \left( \frac{\Delta}{2} \sigma_z \pm \frac{F}{2} \sigma_x\right)  \otimes \mathbbm{1} \\
 & &\pm J ( \sigma_+ \otimes \sigma_-  + \sigma_- \otimes \sigma_+),
\end{eqnarray}
with the following relation between the parameters:
\begin{eqnarray} \label{eq:HoptCondN2}
J=- \Delta \quad \textrm{and} \quad | \Delta | \gg |F| \gg \gamma.
\end{eqnarray}
This means that, strictly speaking, $\rho^{*}_{\Psi_\pm}$ is the stationary state only in the limit of $|\Delta/F| \rightarrow \infty$ and $|F/\gamma| \rightarrow \infty$. However, already for $|\Delta/F| \approx 10$ and $|F/\gamma| \approx 10$, both the concurrence and the $\Psi_\pm$-fidelity of the stationary state reach more than $98\%$ of the optimal value $\frac{1}{2}$.

\begin{figure}[tb]
	\includegraphics[width=.40\textwidth]{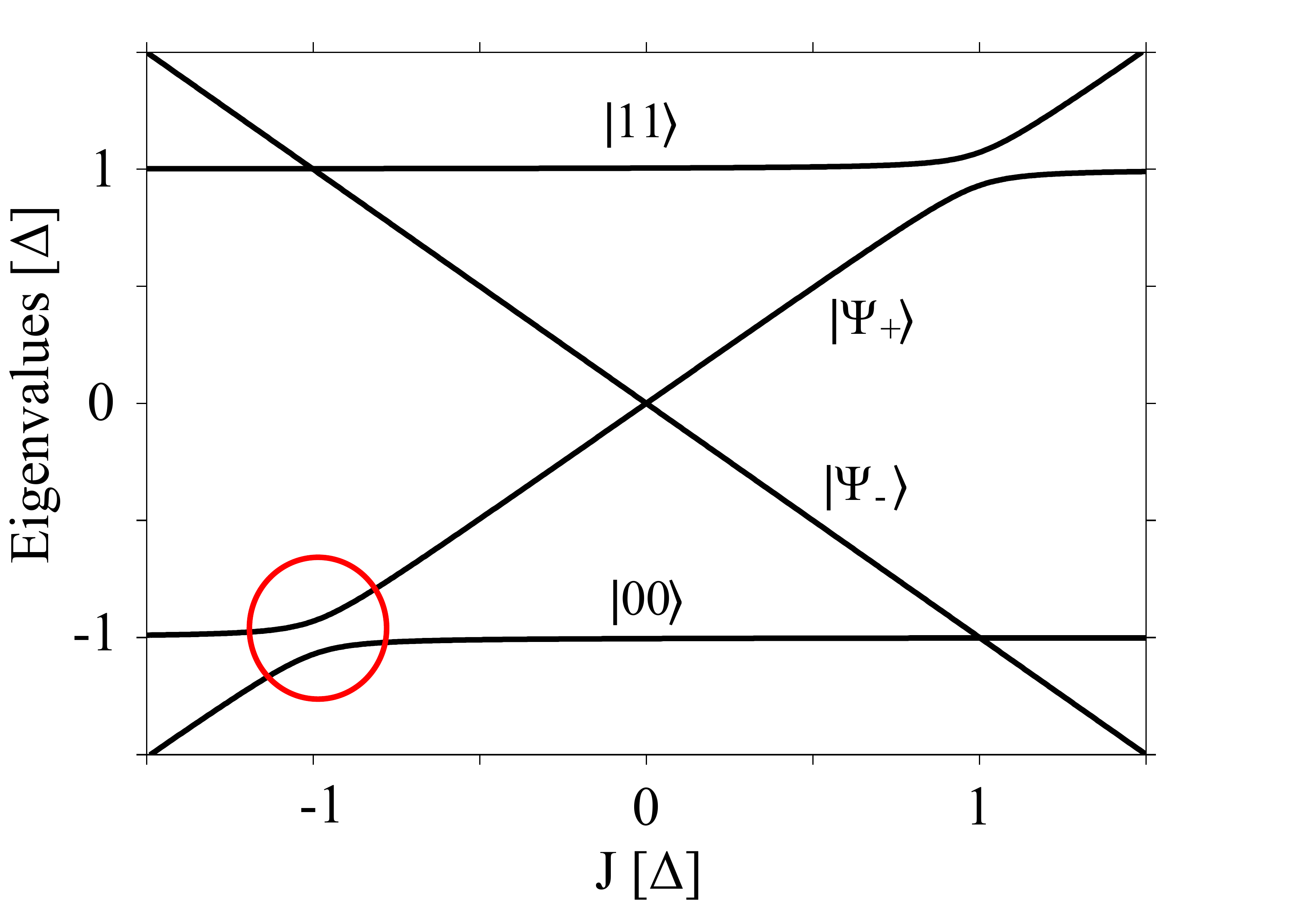}
\caption{(Color online) Spectrum of the Hamiltonian $H^{*}_{+}$, Eq.~\eqref{eq:HoptN2}, as a function of the interaction strength $J$. Both axes are scaled by the interaction parameter $\Delta$. The remaining parameter $F$ is fixed to a small, but non-vanishing value ($F=\Delta/10$), as required in \eqref{eq:HoptCondN2}. (Within the interpretation suggested in Sec. \ref{ssec:ExpRealization}, this corresponds to a weak amplitude of the driving field.) The avoided crossing at $J=- \Delta$ (red circle) establishes $\rho^{*}_{\Psi_+}$ as the stationary state.}\label{fig:spectrumHopt}
\end{figure}

In the following, we discuss in physical terms why $\rho^{*}_{\Psi_+}$ emerges as the stationary state, given the Hamiltonian $H^{*}_{+}$. (The discussion for $\rho^{*}_{\Psi_-} $ and $H^{*}_{-}$ is completely analogous.) Fig.~\ref{fig:spectrumHopt} shows the spectrum of $H^{*}_{+}$ as a function of the interaction strength $J$ in the relevant regime $|\Delta| \gg |F|$. The spectrum can be explained by a perturbative analysis in $F$: At $F=0$, the eigenstates of $H^{*}_{+}$ are $\ket{00}$, $\ket{\Psi_\pm}=\frac{1}{\sqrt 2}(\ket{01}\pm\ket{10})$, and $ \ket{11}$, with corresponding energy levels $-\Delta$, $\pm J$, and $\Delta$. These expressions describe the spectrum in Fig.~\ref{fig:spectrumHopt} already quite well, apart from the fact that they do not explain the two avoided level crossings at $J=\pm \Delta$. To derive the first order correction in $F$, we express the perturbation operator in terms of the unperturbed eigenstates:
\begin{equation}
\sigma_x \otimes \mathbbm{1} + \mathbbm{1}  \otimes  \sigma_x  = \sqrt{2}  \left( \ket{\Psi_+}\bra{11} + \ket{\Psi_+}\bra{00} \right) + h.c.
\end{equation}
Since the perturbation has no diagonal elements in the unperturbed basis, the energy levels are not shifted (to first order). The perturbation alters the spectrum only when the levels $\ket{\Psi_+}$ and $\ket{11}$, or $\ket{\Psi_+}$ and $\ket{00}$, get close to each other. This is the case for $J = \pm \Delta$. There, the perturbation lifts the degeneracy and leads to an avoided crossing of width $F/\sqrt{2}$ between the corresponding levels.

Along with the energy levels, also the eigenstates are modified in the avoided crossing. In fact, in the center of the avoided crossing they become the balanced superposition of the participating levels. For example, at $J = \Delta$, the two states with lower energy, $\ket{00}$ and $\ket{\Psi_-}$, remain unchanged (since they are not coupled by the perturbation, and hence rather cross than anticross), whereas the energetically higher lying states $\ket{11}$ and $\ket{\Psi_+}$ transform into $\frac{1}{\sqrt 2} ( \ket{11} \pm \ket{\Psi_+})$ at the avoided crossing. The same happens at $J = -\Delta$, where the two levels of lower energy anti-cross (see red circle in Fig.~\ref{fig:spectrumHopt}), and the associated eigenstates turn into $\frac{1}{\sqrt 2} ( \ket{00} \pm \ket{\Psi_+})$.

The perturbation-induced transformation of the eigenstates at the avoided crossing $J = -\Delta$ is the underlying mechanism that renders $\rho^{*}_{\Psi_\pm}$ of Eq.~\eqref{eq:rhoopti} the stationary state. This can be explained as follows: As expressed by \eqref{eq:HoptCondN2}, we require the Hamiltonian dynamics to be strong compared to the dissipation rate, $|F|,|\Delta|,|J|\gg \gamma$. In this regime, the right-hand side of the master equation \eqref{eq:Lindbladeq} can only vanish if the Hamiltonian part does so, i.e., if $[H,\rho]=0$. Thus, the stationary state $\rho$ necessarily commutes with the Hamiltonian and therefore becomes diagonal in an eigenbasis $\{\ket{\alpha}\}$ of $H$,
\begin{equation}\label{eq:rhodiagonal}
\rho=\sum_\alpha p_\alpha \ket{\alpha}\bra{\alpha}.
\end{equation}
Albeit comparatively weak, the incoherent part of the master equation is not irrelevant for the stationary state $\rho$, since it determines the weights $p_\alpha$ of the mixture \eqref{eq:rhodiagonal}: Inserting \eqref{eq:rhodiagonal} into \eqref{eq:stationaritycond} leads to the rate equation
\begin{equation}\label{eq:rateequation}
0 = \dot p_\alpha = \sum_\beta ( M_{\beta\alpha} p_\beta - M_{\alpha\beta} p_\alpha),
\end{equation}
with transition rates $M_{\alpha\beta} \equiv \sum_k | \braket{\beta | L_k | \alpha} | ^2$ that describe the probability flow from $\ket{\alpha}$ to $\ket{\beta}$. 
The stationary weights $p_\alpha$ are obtained by extracting the eigenvector with zero eigenvalue of the matrix $P$, which is defined by $P_{\alpha\beta} = M_{\alpha\beta} - \delta_{\alpha\beta} \sum_{\beta'} M_{\alpha\beta'}$ \footnote{
The fact that $P$ has at least one vanishing eigenvalue is ensured by the Perron-Frobenius theorem \cite{Horn:1990}, when applied to the stochastic matrix $P+\mathbbm{1}$.
}. The speed of convergence to the stationary state is then determined by the spectral gap of $P$, i.e., by the second-smallest eigenvalue beyond the stationary eigenvalue zero.

We derived above that the eigenstates of $H^{*}_{+}$ at the avoided crossing $J=-\Delta$ are $\ket{1}\equiv \frac{1}{\sqrt 2} ( \ket{00} + \ket{\Psi_+})$, $\ket{2}\equiv \frac{1}{\sqrt 2} ( \ket{00} - \ket{\Psi_+})$, $\ket{3} \equiv \ket{\Psi_-}$ and $\ket{4} \equiv \ket{11}$. The transition rates read then
\begin{eqnarray}
&M_{11} = M_{22} = M_{12} = M_{21} = \frac{\gamma}{4}, \nonumber\\
& M_{31}=M_{32}=M_{41} = M_{42}= \frac{\gamma}{2},  \quad M_{43}= \gamma,
\end{eqnarray}
while all other rates $M_{\alpha\beta}$ vanish.
This leads to stationary weights $p_{1}=p_{2} =\frac{1}{2}$ and $p_3=p_4=0$. From this, we find the stationary state to be the desired, optimal target state $\rho^{*}_{\Psi_\pm}$:
\begin{eqnarray}
\rho &=& \frac{1}{2} \ket{1}\bra{1} + \frac{1}{2} \ket{2}\bra{2} \nonumber\\
& = & \frac{1}{2} \ket{00}\bra{00} +  \frac{1}{2} \ket{\Psi_+}\bra{\Psi_+} \overset{\eqref{eq:rhoopti}}{=} \rho^{*}_{\Psi_\pm}.
\end{eqnarray}
The speed of convergence to this state is given by $\frac{\gamma}{2}$. This implies that every initial state approaches $\rho^{*}_{\Psi_\pm}$ with precision $\epsilon$ in a finite time of the order of $\gamma^{-1}\log(\epsilon^{-1})$.

As soon as one tunes the parameters away from the center of the avoided crossing, the first and second eigenstates of $H^{*}_{+}$ read $\ket{1'}\equiv\ket{00}$, $\ket{2'} \equiv\ket{\Psi_+}$, while $\ket{3} \equiv \ket{\Psi_-}$ and $\ket{4} \equiv \ket{11}$ remain unchanged. This leads to non-vanishing rates
\begin{equation}
M_{2'1'} = M_{31'}= M_{42'}=  M_{43}= \gamma.
\end{equation}
With this, the rate equation \eqref{eq:rateequation} leads to stationary weights $p_\alpha=\delta_{\alpha,1'}$. Thus, the stationary state is the separable deexcited state $\rho=\ket{00}\bra{00}$; the speed of convergence towards this state is $\gamma$.
For completeness, we mention that the same analysis at the avoided crossing $J=+\Delta$ leads to the stationary state $\rho=\ket{00}\bra{00}$, as well.

\subsection{Experimental realization}\label{ssec:ExpRealization}
The optimal Hamiltonian $H^{*}_{+}$, Eq.~\eqref{eq:HoptN2}, has a surprisingly simple structure and can therefore be implemented in various experimental setups, as discussed in the following.

The first two terms of $H^{*}_{+}$ describe an external field that locally interacts with both qubits. $F$ and $\Delta$, respectively, refer to the field strength in the $x$ and $z$ directions. The third term represents an ``excitation exchange'' interaction of strength $J$ between the qubits. This situation can directly be realized, e.g., with superconducting qubits \cite{Steffen:2006}. 
An alternative, generic implementation of \eqref{eq:HoptN2} that is applicable to almost any experimentally available two qubit system -- be it of quantum optical or solid state nature -- relies on periodic driving: Two qubits with identical level splitting $\omega_0$ are driven by a monochromatic external field of amplitude $F$ and frequency $\omega$, and interact via a 1D Ising interaction of strength $J$. The system Hamiltonian is
\begin{eqnarray}
H(t) &=& \mathbbm{1} \otimes \left(\frac{\omega_0}{2} \sigma_z + F  \cos(\omega t) \sigma_x \right) \nonumber\\
 & & +\left(\frac{\omega_0}{2} \sigma_z + F  \cos(\omega t)  \sigma_x \right) \otimes \mathbbm{1}  + J \ \sigma_x \otimes \sigma_x.
\end{eqnarray}
Performing a rotating frame transformation, it becomes
\begin{eqnarray}
H_\textrm{rf}(t) & =& e^{-i \frac{\omega}{2} ( \sigma_z \otimes \mathbbm{1} + \mathbbm{1} \otimes \sigma_z) } H(t) e^{i \frac{\omega}{2} ( \sigma_z \otimes \mathbbm{1} + \mathbbm{1} \otimes \sigma_z) } \nonumber\\
& = & \mathbbm{1} \otimes \left(\frac{\Delta}{2} \sigma_z  + \frac{F}{2}  (\sigma_x + \sigma_- e^{2i\omega t} + \sigma_+ e^{-2i\omega t}) \right) \nonumber\\
 & + & \left(\frac{\Delta}{2} \sigma_z  + \frac{F}{2}  (\sigma_x + \sigma_- e^{2i\omega t} + \sigma_+ e^{-2i\omega t}) \right) \otimes \mathbbm{1} \nonumber\\
 & +& J \left( \sigma_+ \otimes \sigma_- + \sigma_- \otimes \sigma_+ \right. \nonumber\\
 & &+ \left. e^{2i\omega t} \sigma_- \otimes \sigma_- + e^{-2i\omega t} \sigma_+ \otimes \sigma_+ \right),
\end{eqnarray}
where we have identified the detuning $\omega_0-\omega$ with the parameter $\Delta$. As long as the driving amplitude $F$, the detuning $\Delta$, and the interaction strength $J$ are much smaller than the level splitting $\omega_0$, one can safely neglect the time-dependent parts of $H_\textrm{rf}(t)$ in a rotating wave approximation, leading to $H_\textrm{rf} = H^{*}_{+}$, as desired. Condition \eqref{eq:HoptCondN2} can be met in the experiment by adjusting the frequency $\omega$ of the driving field such that the detuning $\Delta=\omega_0-\omega$ matches $-J$. The driving amplitude $F$ does not have to be tuned to a specific value, but only has to be much weaker than the detuning, and much larger than the rate $\gamma$ of spontaneous decay. In summary, the desired scenario can be implemented by simply driving two interacting qubits at the right frequency \footnote{Note that this is not just due to $H_\textrm{rf} = H^{*}_{+}$, but also relies on the fact that the rotating frame transformation does not alter the dissipator $\myD(\rho)$ for spontaneous decay. Furthermore, it is important that the transformation is local, and therefore does not alter the entanglement properties of $\rho$, which we are interested in here.}. As a side remark, we point out that enhancement of entanglement at avoided crossings is also observed in periodically driven, \textit{closed} quantum systems \cite{Sauer:2012}.

\subsection{Generalization to $N$ qubits}\label{sec:OptimalHN}

The Hamiltonian $H^{*}_{+}$, Eq.~\eqref{eq:HoptN2}, has a natural extension to $N>2$ qubits:
\begin{eqnarray} \label{eq:HoptN}
H^{(N)} & = & \sum_{i=1}^N \left(\frac{\Delta}{2}  \sigma_z^{(i)} + \frac{F}{2}  \sigma_x^{(i)}\right) \nonumber\\
 & &+ \sum_{i<j}^N J ( \sigma_+^{(i)} \sigma_-^{(j)}  + \sigma_-^{(i)} \sigma_+^{(j)}). 
\end{eqnarray}
(The notation $\sigma_z^{(i)}$ refers to a Pauli operator $\sigma_z$ acting on the $i$th qubit.) It can be implemented in complete analogy to the two-qubit scenario discussed in the previous section: Consider $N$ qubits with identical level splitting $\omega_0$, driven by an external field of amplitude $F$ and frequency $\omega$, such that $\Delta$ corresponds to the detuning $\omega_0-\omega$. Every pair of qubits $(i,j)$ interacts via a $\sigma_x^{(i)} \sigma_x^{(j)}$ interaction of equal strength $J$. In rotating wave approximation, such a setup is described by $H^{(N)}$. In the following, we discuss the entanglement of the resulting stationary $N$-qubit state.

Under the combined action of $H^{(N)}$ and spontaneous decay of each qubit with rate $\gamma$, the stationary state $\rho^{(N)}$ is a 50:50 mixture of the deexcited state $\ket{0}^{\otimes N}$ with the $N$ qubit W state $\ket{\textrm{W}_N}\equiv\frac{1}{\sqrt N} (\ket{10\dots 0} + \ket{01\dots 0} + \dots + \ket{0\dots 01})$,
\begin{equation}\label{eq:rhooptiN}
 \rho^{(N)} = \frac{1}{2} \ket{0}^{\otimes N}\bra{0}^{\otimes N} + \frac{1}{2} \ket{\textrm{W}_N}\bra{\textrm{W}_N},
\end{equation}
if the detuning parameter $\Delta$ is adjusted to $J(1-N)$, and if $| \Delta | \gg |F| \gg \gamma$ is fulfilled.
This will be derived below. For $N=2$ qubits, this gives precisely the findings discussed in the previous sections: By choosing the appropriate detuning parameter $\Delta=-J$, the stationary state becomes $\rho^{(2)} \equiv \rho^{*}_{\Psi_+}$, with concurrence $\mathcal{C}(\rho^{*}_{\Psi_+})=\frac{1}{2}$. For $N>2$, we have evaluated the generalized $N$-qubit concurrence \cite{Carvalho:2004,Mintert:2005} of $\rho^{(N)}$ numerically up to $N= 5$ and found that it is always half of the concurrence of the pure W state \footnote{Defining $\rho_x\equiv(1-x) \ket{0}^{\otimes N}\bra{0}^{\otimes N} + x \ket{\textrm{W}_N}\bra{\textrm{W}_N}$, we even observed $\mathcal{C}(\rho_x) = x \, \mathcal{C}(\ket{\textrm{W}_N}\bra{\textrm{W}_N})$ for \textit{all} $x \in [0,1]$, not just for $x=\frac{1}{2}$.}, i.e., $\mathcal{C}(\rho^{(N)})=\frac{1}{2}\mathcal{C}(\ket{\textrm{W}_N})$. The numerical value of $\mathcal{C}(\ket{\textrm{W}_N})$ depends on the normalization one chooses in the definition of the $N$-qubit concurrence. With the convention used in \cite{Mintert:2005}, it is $\mathcal{C}(\ket{\textrm{W}_N}) = \sqrt{2(1-1/N)}$.
Thus, $H^{(N)}$ leads to a substantially entangled stationary state for any number of qubits $N$.

We emphasize that the choice of $H^{(N)}$ in \eqref{eq:HoptN} is heuristically motivated, as a natural extension of the optimal two qubit Hamiltonian \eqref{eq:HoptN2} to $N>2$ qubits. A priori, there is no reason for the resulting stationary state $\rho^{(N)}$ to be optimal with respect to the generalized $N$-qubit concurrence or to the fidelity with respect to a maximally entangled state; in particular so, since there is no unique notion of a maximally entangled state in the multipartite case \cite{Mintert:2005}. However, a statistical analysis of random Hamiltonians for $N=3$, similar to the one presented in Sec. \ref{ssec:MCsampling}, reveals that typical stationary states have poor concurrence values in the range of $0 \dots 0.25$, whereas $\rho^{(3)}$ yields $\mathcal{C}(\rho^{(3)})=\frac{1}{2}\sqrt{4/3}\approx 0.57$. Hence, even more than in the case of two qubits, $H^{(3)}$ yields a stationary state of exceptionally high entanglement.

\begin{figure}[tb]
	\includegraphics[width=.40\textwidth]{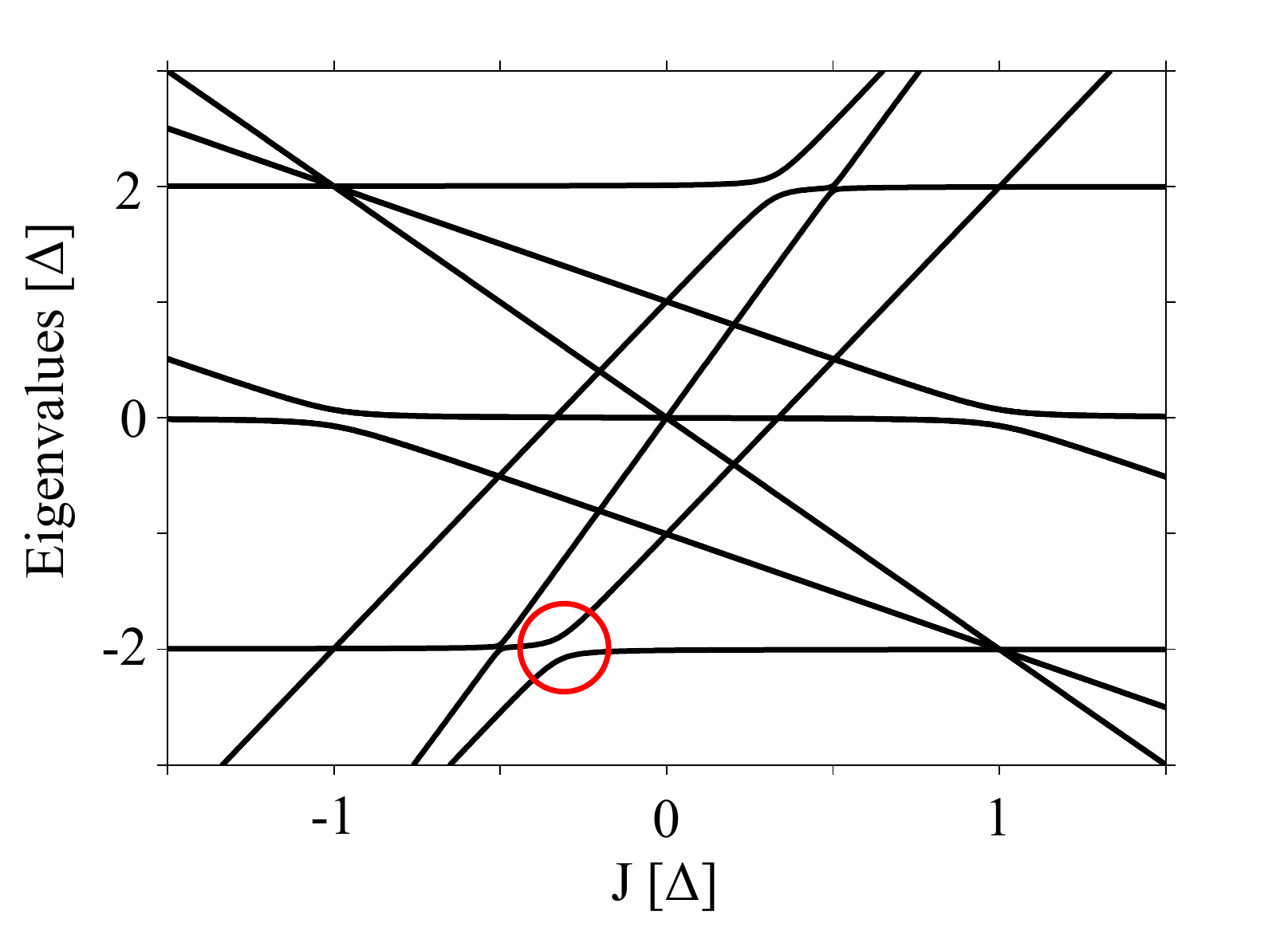}
\caption{(Color online) Spectrum of the many-qubit  Hamiltonian \eqref{eq:HoptN} for $N=4$ qubits. The driving amplitude is $F=\Delta/10$. At the avoided crossing at $J=-\Delta /3$ (red circle), the stationary state turns into the 50:50 mixture \eqref{eq:rhooptiN} of the deexcited state $\ket{0000}$ and the four-qubit W state $\ket{\textrm{W}_4}$.}\label{fig:spectrumHoptN4}
\end{figure}

To confirm that \eqref{eq:rhooptiN} is indeed the stationary state of $H^{(N)}$, we analyze the spectrum of $H^{(N)}$ for $N=4$ qubits in Fig.~\ref{fig:spectrumHoptN4}. Apart from the fact that it involves more levels, it is very similar to its two-qubit counterpart in Fig.~\ref{fig:spectrumHopt}. The energy levels depend linearly on $J$, with different slope. Some levels avoid crossing, while others cross exactly. To understand the spectrum in detail, it is convenient to introduce the collective spin operator $\vec S = \sum_i \frac{1}{2} {\vec\sigma}^{(i)}$, which formally corresponds to the angular momentum operator of a spin-$\frac{N}{2}$ system. In this notation, our Hamiltonian \eqref{eq:rhooptiN} reads
\begin{eqnarray} \label{eq:HoptNalt}
H^{(N)} = \Delta \cdot S_z + F \cdot S_x + J \left( {\vec S}^2 - S_z^2 - \frac{N }{2} \right).
\end{eqnarray}
At $F=0$, $H^{(N)}$ contains only ${\vec S}^2$ and $S_z$, and its eigenstates are therefore the well-known angular momentum eigenstates $\ket{l,m}$ \cite{Sakurai:1993}, with $l=0,1,\dots,\frac{N}{2}$, and $m=-l,\dots,l$. (We assume $N$ to be even here, but the case of odd $N$ is completely analogous). The corresponding energy eigenvalues are
\begin{eqnarray} \label{eq:HoptNspec}
E_{lm}  = m\cdot \Delta + J \left( l(l+1) - m^2 - \frac{N}{2} \right).
\end{eqnarray}
This explains the linear dependence of the eigenvalues on $J$. To understand the anti-crossings, we employ again first order perturbation theory in the driving strength $F$. The perturbation operator is $S_x = (S_+ + S_-)/2$, and its matrix elements in the unperturbed basis $\ket{l,m}$ are \cite{Sakurai:1993}
 \begin{eqnarray}
 \braket{l',m' | (S_+ + S_-) | l,m} = \nonumber\\
\delta_{ll'} (\sqrt{(l-m)(l+m+1)} \delta_{m',m+1} \nonumber\\
+\sqrt{(l+m)(l-m+1)} \delta_{m',m-1}).
 \end{eqnarray}
Hence, only levels with the same quantum number $l$ and neighboring $m$ interact (at first order). For our purposes, the avoided crossing between $\ket{\frac{N}{2},-\frac{N}{2}}$ and $\ket{\frac{N}{2},-\frac{N}{2}+1}$ is most interesting. $\ket{\frac{N}{2},-\frac{N}{2}}$ is simply the deexcited state $\ket{0}^{\otimes N}$, and $\ket{\frac{N}{2},-\frac{N}{2}+1}$ is the $N$ qubit W state $\ket{\textrm{W}_N}$.
According to \eqref{eq:HoptNspec}, both states come close in energy at $J = \Delta/(1-N)$, as marked by the circle in Fig.~\ref{fig:spectrumHoptN4}.
At the center of the resulting anti-crossing, the balanced superpositions $\frac{1}{\sqrt 2}(\ket{0}^{\otimes N} \pm \ket{\textrm{W}_N})$ become eigenstates of $H^{(N)}$.

To determine the stationary state in the regime of $| \Delta | \gg |F| \gg \gamma$, one can proceed in complete analogy with the case of two qubits. As long as $J$ is chosen different from position of the anti-crossing at $J = \Delta / (1-N)$, the analysis yields the de-excited state $\rho^{(N)}=\ket{0}^{\otimes N}\bra{0}^{\otimes N}$. At the avoided crossing, however, the eigenstates are transformed, and the rate equation \eqref{eq:rateequation} leads to different stationary weights. This results in the stationary state $\rho^{(N)}$ of Eq.~\eqref{eq:rhooptiN}.

In summary, $N$ qubits undergoing spontaneous decay at rate $\gamma$ can be prepared in the highly entangled state \eqref{eq:rhooptiN} in the following way: Implement the Hamiltonian \eqref{eq:HoptN} as described, tune it into the regime of $| \Delta | \gg |F| \gg \gamma$, adjust the detuning parameter $\Delta$ to $J(1-N)$, and wait for the system to reach its stationary state.

\section{Conclusion}\label{sec:Conclusion}

In the present work, we have investigated the entanglement properties of the stationary states of the Lindblad master equation \eqref{eq:Lindbladeq}. Our aim was to investigate to what extent robust entangled states can be prepared in the presence of naturally occurring incoherent processes with the generic local structure \eqref{eq:locL}.

As a first, general result, we found that pure stationary states of $N$ qubits are necessarily separable with respect to any subsystem which is subject to a local incoherent process. Hence, stationary states with a finite amount of entanglement are necessarily mixed in this general scenario.

For two qubits undergoing spontaneous decay, we found that typical stationary states exhibit limited entanglement, as quantified by either the concurrence or the Bell state fidelities.
The \emph{most entangled} among all conceivable stationary states of two qubits under spontaneous decay, on the other hand, was shown to have exceptionally high entanglement $\mathcal{C}(\rho^{*}_{\Psi_\pm})=\frac{1}{2}$, a value that is not found in the statistical ensemble of typical stationary states. In Sec.~\ref{sec:OptimalH}, we discussed in detail the Hamiltonian $H_\pm$ that yields this optimal stationary state, proposed concepts for its experimental implementation, and found that its generalization to $N$ qubits yields a stationary state with a substantial amount of multi-partite entanglement.

\begin{acknowledgments}
We thank Ugo Marzolino for fruitful discussions. S.S. acknowledges financial support by the German National Academic Foundation. A.B. acknowledges partial support through COST action MP1006 and by DFG.
\end{acknowledgments}

\appendix
\section{Proofs}\label{sec:AppendixProofs}

\subsection{Separability of target states when both dissipator and Hamiltonian act locally}\label{ssec:AppendixProof1}
\textit{Suppose that $\rho$ is the unique stationary state of the master equation \eqref{eq:Lindbladeq}, with Hamiltonian $H$ and local Lindblad operators $L_{k}$, as defined in \eqref{eq:locL}. If $\rho$ is entangled, then $H$ must contain nonlocal terms.}
\paragraph*{Proof:}
Suppose that $H$ contains only local terms. Since the $L_{k}$ are local by assumption, the master equation \eqref{eq:Lindbladeq} decomposes into individual evolution equations for each subsystem. Therefore, the product $\rho_1\otimes \dots \otimes \rho_N$ of stationary states of the individual subsystems is a stationary state of the composite system. Since the stationary state of the composite system is unique by assumption, this contradicts the premise of $\rho$ being entangled. Hence, $H$ must contain non-local terms. $\blacksquare$

\subsection{Nonexistence of pure, entangled target states in the presence of local dissipation}\label{ssec:AppendixProof2}

\propeins

\paragraph*{Proof:} Without loss of generality, we assume $k=1$, i.e., the local Lindblad operator $L_1$ acts on the first qubit:
\begin{eqnarray}\label{eq:Ldef}
L_{1} = l\otimes\mathbbm{1}\otimes\dots\otimes\mathbbm{1} .
\end{eqnarray}
Furthermore, $L_1$ (and therefore also $l$) can be assumed traceless: If $\mathrm{Tr}\,L_1\ne 0$, let $L_1'\equiv L_1- c\mathbbm{1}$, and $H'=H + \frac{i}{2} c^* L_1 - \frac{i}{2} c L_1^\dagger$, with $c= (\mathrm{Tr}\,L_1)/(2^N)$. The master equation \eqref{eq:Lindbladeq} is invariant under this transformation, and therefore the above statement also holds for the traceless Lindblad operator $L_1'$, which inherits from $L_1$ the property of being local.

Since the stationary state $\rho=\ket{\psi}\bra\psi$ is pure, it must be an eigenstate of all Lindblad operators of the process. (See Theorem 1 in Ref. \cite{Kraus:2008}, or Proposition 4 in Ref. \cite{Schirmer:2010}.) Hence, we have
\begin{eqnarray}\label{eq:Leigval}
L_1\ket\psi=\alpha\ket\psi
\end{eqnarray}
for some eigenvalue $\alpha\in\mathbb{C}$.
Next, we write down the Schmidt decomposition of $\ket{\psi}$ with respect to the bipartition $\{1\} | \{2\dots N\}$:
\begin{eqnarray}\label{eq:Schmidtdec}
\ket\psi = \lambda_a \ket{a_1}\otimes \ket{a_{2\dots N}} + \lambda_b \ket{b_1}\otimes \ket{b_{2\dots N}}.
\end{eqnarray}
Using \eqref{eq:Ldef} and \eqref{eq:Leigval}, we then have
\begin{eqnarray}
& \lambda_a (l \ket{a_1})\otimes \ket{a_{2\dots N}} + \lambda_b (l \ket{b_1})\otimes \ket{b_{2\dots N}}  \nonumber \\
 = & \lambda_a (\alpha \ket{a_1})\otimes \ket{a_{2\dots N}} + \lambda_b (\alpha \ket{b_1})\otimes \ket{b_{2\dots N}}.
\end{eqnarray}
Assume $\ket\psi$ is \textit{not} separable, i.e., both Schmidt coefficients $\lambda_a$ and $\lambda_b$ are non-zero. Then, it follows that $l \ket{a_1} =\alpha \ket{a_1}$ and $l \ket{b_1} =\alpha \ket{b_1}$, since the Schmidt decomposition ensures that $\ket{a_{2\dots N}}$ and $\ket{b_{2\dots N}}$ are orthogonal. Hence, $l$ has the two-fold degenerate eigenvalue $\alpha$. Since $l$ is a traceless single qubit operator, we have $\textrm{Tr}\,l=2\alpha = 0 \Rightarrow \alpha=0$. Hence, $l$ must be the null operator $l=0$, implying, in turn, $L_1=0$. This trivial case is of course excluded in our premise, and therefore $\ket\psi$ must be separable with respect to the bipartition $\{1\} | \{2\dots N\}$. $\blacksquare$

\section{Example of a pure, unique, entangled stationary state under local dissipation}\label{sec:AppendixExamplePSIent}

\begin{figure}[tb]
	\includegraphics[width=.48\textwidth]{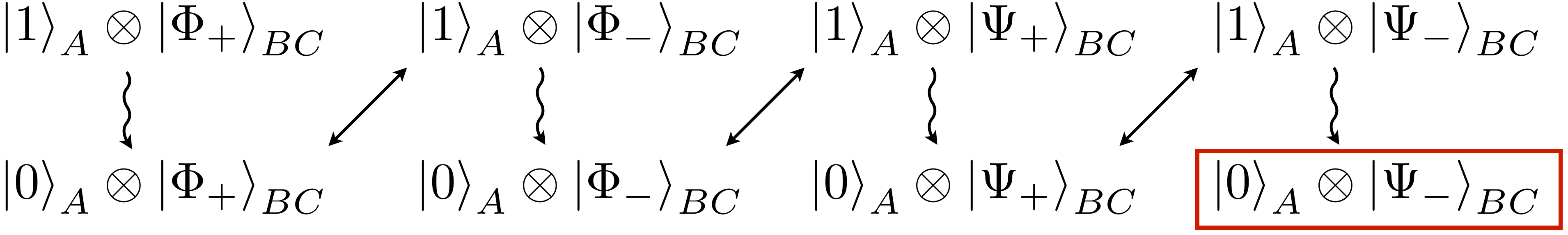}
\caption{(Color online) Example of a local dissipative state preparation scheme for three qubits. Spontaneous decay of qubit $A$ (wavy arrows), together with Hamiltonian \eqref{eq:exampleH} (straight arrows), drives any initial state into the unique, stationary state $\ket{0}_A\otimes\ket{\Psi_-}_{BC}$, in which $B$ and $C$ are maximally entangled.}\label{fig:exampleH}
\end{figure}

To illustrate that even local dissipation can lead to an entangled target state if the Hamiltonian is adequately chosen, we consider the example of three qubits $A$, $B$, and $C$, with the following Hamiltonian:
\begin{eqnarray}\label{eq:exampleH}
H =& & \ket{0}_A\otimes\ket{\Phi_+}_{BC} \bra{1}_A \otimes\bra{\Phi_-}_{BC} \nonumber\\
&+ & \ket{0}_A \otimes\ket{\Phi_-}_{BC} \bra{1}_A \otimes\bra{\Psi_+}_{BC} \nonumber\\
&+ & \ket{0}_A \otimes\ket{\Psi_+}_{BC} \bra{1}_A \otimes\bra{\Psi_-}_{BC} + h.c. \quad.
\end{eqnarray}
$\ket{\Phi_\pm}_{BC}$ and $\ket{\Psi_\pm}_{BC}$ denote the maximally entangled Bell states, shared between qubit $B$ and $C$, as defined in Eqs.~\eqref{eq:BelldefPhi} and \eqref{eq:BelldefPsi}.
For the dissipative dynamics, we assume spontaneous decay of qubit $A$ only, i.e., $L_1= \sigma_-^{(1)}$. As immediately apparent from Fig.~\ref{fig:exampleH}, the unique stationary state of this system is the pure state $\ket{0}_A\otimes\ket{\Psi_-}_{BC}$, in which $B$ and $C$ are maximally entangled.
We emphasize, however, that this example is not quite generic, since it relies on the fact that only qubit $A$ (which serves as a kind of ancillary system here) undergoes an incoherent process, and that the qubits are coupled in a peculiar way via the Hamiltonian $H$.
If one aims for genuine $3$-partite entanglement in this system, the stationary state can no longer be pure, but it is necessarily mixed, as proven in Appendix~\ref{sec:AppendixProofs}.\ref{ssec:AppendixProof2}.

\section{Condition (\ref{eq:conservespectrum}) for two qubits in generalized Bloch notation}\label{sec:AppendixQuadricN2}

In the following, we transform condition \eqref{eq:conservespectrum} for $n=2$ into the generalized Bloch notation for two qubits, with the specific choice of the Lindblad operators $L_1= \sigma_-\otimes \mathbbm{1}$ and $L_2= \mathbbm{1}\otimes \sigma_-$ which describe spontaneous decay of each qubit (at rate $\gamma$).

The generalized Bloch vector $\vec{\mathfrak{r}}_\rho$ is defined via $(\vec{\mathfrak{r}}_\rho)_{4i+j}=\textrm{Tr}[(\sigma_i \otimes \sigma_j) \  \rho]$, with $\sigma_0 \equiv \mathbbm{1}_2, \sigma_1 \equiv \sigma_x, \sigma_2 \equiv \sigma_y, \sigma_3 \equiv \sigma_z$ (i.e., the indices $i$ and $j$ run from $0$ to $3$). Since $(\vec{\mathfrak{r}}_\rho)_{0}=\textrm{Tr}(\rho)=1$, one only has to consider the remaining 15 components \footnote{As a side remark, we point out that state space $\mathcal{Q}$ is no longer a ball in this case \cite{Bengtsson:2006}, in contrast to the Bloch representation of a single qubit.}. Inserting this definition into condition \eqref{eq:conservespectrum} for $n=2$, one obtains a quadratic expression in the Bloch vector:
\begin{equation}
\vec{\mathfrak{r}}_\rho \cdot( \mathfrak{D}\vec{\mathfrak{r}}_\rho +  \vec{\mathfrak{c}}) = 0.
\end{equation}
The $15 \times 15$ matrix $\mathfrak{D}$ has diagonal entries $\textrm{diag}(\mathfrak{D})= -\frac{\gamma}{2} \ (1,1,2,1,2,2,3,1,2,2,3,2,3,3,8)$, and its non-zero off-diagonal elements are $\mathfrak{D}_{1,7}=\mathfrak{D}_{2,11}=\mathfrak{D}_{3,15}=\mathfrak{D}_{4,13}=\mathfrak{D}_{8,14}=\mathfrak{D}_{12,15}=-\gamma$. The elements of the vector $\vec{\mathfrak{c}}$ are zero, except for $(\vec{\mathfrak{c}})_{3}=(\vec{\mathfrak{c}})_{12}=-\gamma$.


\end{document}